\documentclass[10pt,a4paper]{article}
\usepackage[utf8]{inputenc}
\usepackage{amsmath}
\usepackage{amsfonts}
\usepackage{amssymb}
\include{Qcircuit}
\usepackage{graphicx}
\usepackage[utf8]{inputenc}
\usepackage[T1]{fontenc}
\usepackage{geometry}
\usepackage{amsmath}
\usepackage{array}

\usepackage{authblk}
\DeclareGraphicsExtensions{.png}
\usepackage{cite}

\begin{document}
\title{Rooted-tree network for optimal non-local gate implementation}

\author[1]{Nilesh Vyas$^{\footnote{nilesh.vyas@niser.ac.in}}$}
\author[2]{Debashis Saha$^{\footnote{debashis.saha@phdstud.ug.edu.pl}}$}
\author[3]{Prasanta K. Panigrahi$^{\footnote{pprasanta@iiserkol.ac.in}}$}

\affil[1]{\it National Institute of Science Education and Research, Institute of Physics Campus, Sachivalaya Marg, PO: Sainik School, Bhubaneswar - 751005, India}
\affil[2]{\it Institute of Theoretical Physics and Astrophysics, University of Gda{\'{n}}sk, Gda{\'{n}}sk - 80952, Poland}
\affil[3]{\it Indian Institute of Science Education and Research Kolkata, Mohanpur - 741246, India }

\maketitle

\begin{abstract}

A general quantum network for implementing non-local control-unitary gates, between remote parties at minimal entanglement cost, is shown to be a $rooted-tree$ structure. Starting from a five party scenario, we demonstrate the local implementation of simultaneous control-Hermitian and multiparty control-unitary gates in an arbitrary $n$-party network. Previously established networks are shown to be special cases of this general construct.
\end{abstract}

 \section {Introduction}
Remote implementation of quantum gates is crucial for distributed quantum computation, as it enables one to non-locally design useful quantum circuits.
Non local implementation of local quantum gate involves multiparty operations, making essential use of entanglement, local operation and classical communication (LOCC). Teleportation, dense coding and a host of other communication protocols are also usefully employed. Designing a quantum network and implementing gate teleportation with minimal entanglement cost have been of deep interest to both theoretical and experimental community. Eisert {\it et.al.} \cite{1} proposed a method for optimal local implementation of nonlocal quantum gates. Experimental teleportation of the quantum controlled-NOT gate was realized by Bouwmeester {\it et.al.} \cite{3} and Huang {\it et.al.} \cite{4}. Subsequently,  number of works investigated efficient implementation of these gates, as well as entanglement requirements in bipartite \cite{5,6} and multipartite \cite{7} scenarios. The communication complexity of implementing  remote operations \cite{8} and teleportation of nonlocal quantum gates with two \cite{9} control  agents have been reported. \\
The proposal of Eisert {\it et.al.} involved a network in a {\it parallel} configuration, where each party shares a pair of entangled states with the target party and no entangled state is shared between control parties. The target party, therefore, has to deal with large number of entangled pairs, making this protocol inherently complicated to realize. To avoid this problem, another configuration has been considered by \cite{10} which uses a linear entangled channel that can be visualized as a {\it series} connection. Here each control party shares entanglement with the adjacent one and only one of them is entangled with the target party. As each party deals with only two entangled pairs, the series network has an advantage over the parallel one, although the classical communication cost increases significantly.\\
In this paper we present a generalized network for gate teleportation, using a branched tree network channel, known as {\it rooted-tree} network in graph theory parlance. This network enables us to implement a multiparty gate, according to the capacity of the entangled channels shared by each remote party, simultaneously reducing the classical communication cost, as compared to the linear network. The above mentioned two networks are found to be the special cases of the rooted-tree structure. We explicate the proposed protocol by implementing a controlled-Hermitian $(C\mathcal{H})$ gate and multiparty controlled-unitary $(CU)$  gate, a generalized form of the Toffoli gate.\\ 
In the following section, we establish the rooted-tree network as the most general entanglement distribution structure amongst remote parties, which makes optimal use of entanglement resources. In the subsequent two sections, the protocol for implementation of controlled-Hermitian gate and multiparty controlled unitary gate are carried out. For explicitness we demonstrate the protocol in five party system, with four control parties and one target party which is then generalize for arbitrary control parties. The required entanglement and classical communication resources, as well as the number of implementation steps required for optimal gate implementation are enumerated. Finally, we conclude after pointing out several directions for future research.

\section {Rooted tree network}
As mentioned earlier quantum computation in a distributed scenario, as also implementation of desired tasks at remote locations, requires implementation of non-local quantum control gates. The parallel network protocol is inherently complicated to realize and the series network escalates the classical communication cost. Keeping these in mind we design a generalized network optimizing both costs. Here, we take recourse to graph theory for designing a generalized network for optimizing classical communication cost, while keeping the network easy for implementation. In our proposed construct, there are ($n-1$) remote control parties who want to implement control unitary gates on one target party. In the graph theory framework, each remote party is represented by a vertex. Each vertex may have two or more qubits. One or more qubits of a vertex are connected to the qubits of other vertices through edges, made up of Bell states. Since each remote agent is connected to at least one other, the graph representing the quantum network is {\it connected}. A graph is connected, if all the vertices are joined to each other by a path through edges. To ensure an optimal entanglement cost, the number of edges of the relevant graph should be ($n-1)$. It can be shown that a connected graph of $n$ vertices with ($n-1$) edges has to be a {\it tree}. For this to occur, the graph should be devoid of  any circuit, which connects a vertex with itself through edges in a closed path. It is evident that, if there exists a circuit in the network, then by removing one edge one can still make a connected graph. However, the fact that a connected graph of $n$ vertices has at least ($n-1)$ edges, contradicts that a connected graph with ($n-1)$ edges has circuit \cite{27}. 

In the present case with one target party and $(n-1)$ control parties, one can define a direction for each edge. The network has a designated vertex called the root, representing the target party. Vertices representing the control parties are leaves having edges directed away from the root. This general entanglement distribution between remote parties forms a rooted-tree structure, as illustrated  in  Fig.1.
\begin{figure*}[h]
   \centering
  \includegraphics[scale=0.8]{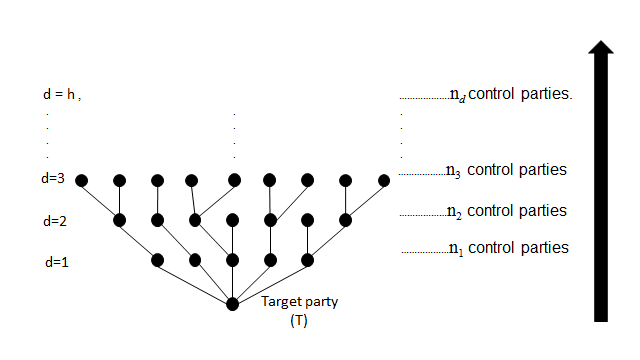}
    \caption{ Rooted tree network with the target party, $T$, as the root node, with the rest of the control parties arranged in a pyramidal structure in varying depths, $d$. The upward arrow on the right shows the direction of each edge, directed away towards control parties starting from the root.}
\end{figure*}

One defines the depth ($d$) of a control party as the number of edges present in the unique path from the target party. The height ($h$) of the rooted tree is the largest depth of the tree and the control party having an outward edge is called a leaf. Here, the number of control parties of the same depth is denoted as $n_{d}$: 
\begin{equation}
n = \sum_{d=1}^h n_{d} + 1,
\end{equation}
where $n$ is the total number of parties.

The rooted tree network is the most general network structure under the condition of optimal entanglement cost, which is $(n-1)$ ebits for $n$ remote parties. We now proceed to the details of the protocol of implementing non-local gates.

 \section {Simultaneous implementation of control-Hermitian gate}
The simultaneous implementation of control-Hermitian gate is a realization of a number of control-Hermitian gates on a common target qubit in a quantum network. If the quantum state $|\psi \rangle_{12...n}$, connects  $n$ remote parties, possessing a single qubit each, with the $n$-th qubit as the target, then the desired state with the gate teleportation protocol will be $C\mathcal{H}^n_1 C\mathcal{H}^n_2...C\mathcal{H}^n_{n-1} |\psi \rangle_{12...n}$. Here, $C\mathcal{H}^j_i$ denotes the control-Hertimian gate where, $i$ is the control qubit and $j$ the target. Although the protocol does not allow one to simultaneously implement a control-unitary gate, it is worth mentioning that many of important gates in quantum computation belong to this category.

For the sake of explicitness, we first consider a five party scenario, maintaining the entangled channel in a rooted-tree pattern, as depicted in Fig.2. 
Let $S_{11}$, $S_{12}$, $S_{21}$ and $S_{22}$ be the four control parties (each control party is denoted as $S_{dj}$, where $d$ represents the depth and $j$ the location in that depth), wanting to simultaneously implement a controlled-Hermitian gate on a target party ($T$).\\
\begin{figure*}[h]
   \centering
  \includegraphics[scale=0.8]{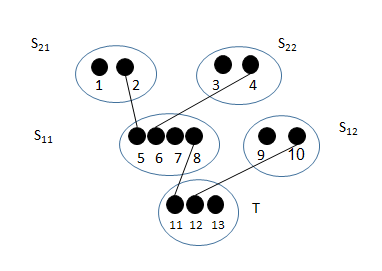}
    \caption{A rooted tree network comprising of four control parties, $S_{dj}$ (where $d$ represents the depth and $j$ the location in that depth), and the target party $T$. In the figure, each party is represented by a system of qubits enclosed in the shape. Numbers as shown below each qubit signifies the qubit possessed by that party.   }
\end{figure*}\\

Let the five parties possess qubits 1, 3, 7, 9, 13 of an arbitrary state 
$ | \psi \rangle_{1, 3, 7, 9, 13} $. 
$S_{21}$ and $S_{22}$ both share Bell states $| \Phi \rangle_{2,5}$ and $ | \Phi \rangle_{4,6}$, respectively with $S_{11}$. $S_{11}$ and $S_{12}$ share Bell states $ | \Phi \rangle_{8,11}$ and $| \Phi \rangle_{10,12}$ respectively with $T$, such that
\begin{equation}
| \Phi \rangle_{2,5} = | \Phi \rangle_{4,6} = | \Phi \rangle_{8,11} = | \Phi \rangle_{10,12}
= \frac{1}{\sqrt{2}}( | 00 \rangle + | 11 \rangle )
\end{equation}

The protocol to implement simultaneous controlled-Hermitian gate, describing each implementation {\it step} is given below. Here, the implementation step is considered as a successive operation, which may be a quantum measurement or some other quantum operation or classical communication, as required. The following steps enumerate the implementation of protocol. \\

{\bf Step 1:} All the leaves {\it i.e.}, $S_{12}$, $S_{21}$ and $S_{22}$ first apply Control-Not ($CN^j_i$, $i$ is the control and $j$ is the target) gate on their respective qubits $CN^{2}_{1}$, $CN^{4}_{3}$, $CN^{10}_{9}$.\\

{\bf Step 2:} In the second step, both $S_{21}$ and $S_{22}$ measure their qubits 2 and 4 in the computational basis.\\

{\bf Step 3:} They then convey the outcomes of the measurements to $S_{11}$ using classical resources. \\

{\bf Step 4:} Depending on the outcome, $S_{11}$ performs the local operations as listed in Table \ref{table 1}, where $\sigma_x^i, \sigma_z^i, \sigma_y^i$ are Pauli operators, with superscript `$i$' indicating the operating qubit.\\
 
\begin{table}[h]
\begin{center}
\begin{tabular}{|c|c|}
\hline 
{\bf Outcome of measurements}& {\bf Local operation after measurements} \\ 
\hline 
$ | 0 \rangle_{2} | 0 \rangle_{4}$ & $CN^8_{5} ~ CN^8_{6}~ CN^8_{7}$ \\ 
\hline 
$| 1 \rangle_{2} | 0 \rangle_{4}$ & $CN^8_{5} ~ CN^8_{6}~ CN^8_{7}  ~ \sigma^{5}_x$ \\ 
\hline 
$| 0 \rangle_{2} | 1 \rangle_{4}$ & $CN^8_{5} ~ CN^8_{6}~ CN^8_{7}  ~ \sigma^{6}_x$ \\ 
\hline
$| 1 \rangle_{2} | 1 \rangle_{4}$ & $CN^8_{5} ~ CN^8_{6}~ CN^8_{7}  ~ \sigma^{5}_x  ~ \sigma^{6}_x $ \\
\hline 
\end{tabular}
\caption{Local operations performed by $S_{11}$ after the outcome of measurement of qubit 2 and 4 by $S_{21}$ and $S_{22}$ respectively }
\label{table 1}
\end{center}
\end{table}

{\bf Step 5:}  $S_{11}$ and $S_{12}$ measure their qubits 8 and 10 in the computational basis.\\

{\bf Step 6:} Next, they convey the outcomes to target party $T$.\\

{\bf Step 7:} Based on their outcome $T$ then performs the local operations as described in Table \ref{table 2},\\
\begin{table}[h]
\begin{center}
\begin{tabular}{|c|c|}
\hline 
{\bf Outcome of measurements}& {\bf Local operation after measurements} \\ 
\hline 
$ | 0 \rangle_{8} | 0 \rangle_{10}$ & $C\mathcal{H}^{13}_{11} ~ C\mathcal{H}^{13}_{12} $ \\ 
\hline 
$| 1 \rangle_{8} | 0 \rangle_{10}$ & $C\mathcal{H}^{13}_{11} ~ C\mathcal{H}^{13}_{12}  ~ \sigma^{11}_x$ \\ 
\hline 
$| 0 \rangle_{8} | 1 \rangle_{10}$ & $C\mathcal{H}^{13}_{11} ~ C\mathcal{H}^{13}_{12}  ~ \sigma^{12}_x$ \\ 
\hline
$| 1 \rangle_{8} | 1 \rangle_{10}$ & $C\mathcal{H}^{13}_{11} ~ C\mathcal{H}^{13}_{12}  ~ \sigma^{11}_x  ~ \sigma^{12}_x $ \\
\hline 
\end{tabular}
\caption{Local operations performed by Target party \textit{T}, based on the outcomes of qubits 8 and 10 measured by $S_{11}$ and $S_{12}$ respectively}
\label{table 2}
\end{center}
\end{table}

{\bf Step 8:} Subsequently, qubits 5 and 6 are measured in the Hadamard basis by $S_{11}$ and qubits 11 and 12 by $T$.\\

{\bf Step 9:} $T$ conveys the measurement outcomes of qubit 11 to $S_{11}$, $S_{21}$, $S_{22}$; and qubit 12 to $S_{12}$. $S_{11}$ conveys the outcomes of qubits 5 and 6 to $S_{21}$ and $S_{22}$ respectively.\\

{\bf Step 10:}  Correspondingly the control parties perform following unitary operations to achieve the desired state shared by the parties,\\

\begin{table}[h]
\begin{center}
\begin{tabular}{|c|c|}
\hline 
{\bf Outcome of measurements}& {\bf Local operation after measurements} \\ 
\hline 
 $ | ++++ \rangle_{5,6,11,12}$ & $I$ \\ 
\hline 
$ | +++- \rangle_{5,6,11,12}$ &  $\sigma^{9}_z$  \\ 
\hline 
$ | ++-+ \rangle_{5,6,11,12}$ & $\sigma^{1}_z$ $\sigma^{3}_z$ $\sigma^{7}_z$ \\ 
\hline 
$ | +-++ \rangle_{5,6,11,12}$ & $\sigma^{3}_z$ \\ 
\hline 
$ | -+++ \rangle_{5,6,11,12}$ & $\sigma^{1}_z$ \\ 
\hline 
$ | ++-- \rangle_{5,6,11,12}$ & $\sigma^{1}_z$ $\sigma^{3}_z$ $\sigma^{7}_z$ $\sigma^{9}_z$ \\ 
\hline 
$ | --++ \rangle_{5,6,11,12}$ & $\sigma^{1}_z$ $\sigma^{3}_z$ \\ 
\hline 
$ | +--+ \rangle_{5,6,11,12}$ & $\sigma^{1}_z$ $\sigma^{7}_z$ \\ 
\hline 
$ | -++- \rangle_{5,6,11,12}$ & $\sigma^{1}_z$ $\sigma^{9}_z$ \\ 
\hline 
$ | +-+- \rangle_{5,6,11,12}$ & $\sigma^{3}_z$ $\sigma^{9}_z$ \\ 
\hline 
$ | -+-+ \rangle_{5,6,11,12}$ & $\sigma^{3}_z$ $\sigma^{7}_z$ \\ 
\hline 
$ | +--- \rangle_{5,6,11,12}$ & $\sigma^{1}_z$ $\sigma^{7}_z$ $\sigma^{9}_z$ \\ 
\hline 
$ | -+-- \rangle_{5,6,11,12}$ & $\sigma^{3}_z$ $\sigma^{7}_z$ $\sigma^{9}_z$ \\ 
\hline 
$ | --+- \rangle_{5,6,11,12}$ & $\sigma^{1}_z$ $\sigma^{3}_z$ $\sigma^{9}_z$ \\ 
\hline 
$ | ---+ \rangle_{5,6,11,12}$ & $\sigma^{7}_z$ \\ 
\hline 
$ | ---- \rangle_{5,6,11,12}$ & $\sigma^{7}_z$ $\sigma^{9}_z$ \\ 
\hline 
\end{tabular}
\caption{Unitary operations performed by control parties to achieve the desired state}
\label{table 3}
\end{center}
\end{table}

The quantum circuit for the simultaneous implementation of $ C\mathcal{H}$ gate as described by the above protocol is depicted in Fig.3. The entanglement resources (ebits) and classical communication resources (cbits) required to obtain the desired state are 4 and 10, respectively. \\

\begin{figure*}[h]
   \centering
 \Qcircuit @C=1em @R=.7em
{
& \lstick{(S_{21})} & \lstick{1} & \qw & \ctrl{1} & \qw & \qw & \qw & \qw & \qw & \qw  & \qw & \qw & \qw & \gate{\sigma_{z}} & \qw & \qw & \gate{\sigma_{z}} & \qw & \qw & \qw  \\
& \lstick{(S_{21})} & \lstick{2} & \ex[3] & \targ & \meter &  &  &  &  &  & &  &  &  \cwx &  &  &  \cwx &   &  &    \\
& \lstick{(S_{22})} & \lstick{3} & \qw & \ctrl{1} & \qw \cwx & \qw & \qw & \qw & \qw & \qw & \qw & \qw & \qw  & \qw \cwx & \gate{\sigma_{z}} & \gate{\sigma_{z}} & \cwx \qw & \qw  & \qw  & \qw \\
& \lstick{(S_{22})} & \lstick{4} \ex[2] & \qw  & \targ & \qw \cwx & \meter &  &  &  &  &  & & &  \cwx &  \cwx & \cwx &  \cwx &  &  &   \\
& \lstick{(S_{11})} & \lstick{5} & ~ & \qw & \targ \cwx & \qw \cwx & \qw & \ctrl{3} & \qw & \qw & \qw & \qw & \gate{H} & \meter \cwx &  \cwx & \cwx &  \cwx &   &    & \\
& \lstick{(S_{11})} & \lstick{6} & \qw & \qw & \qw & \targ \cwx & \qw & \qw & \qw & \ctrl{2} & \qw & \qw & \qw & \gate{H} & \meter \cwx & \cwx &  \cwx &  &   &  \\
& \lstick{(S_{11})} & \lstick{7} & \qw & \qw & \qw & \qw & \qw & \qw & \ctrl{1} & \qw & \qw & \qw & \qw  & \qw & \gate{\sigma_{Z}} & \qw \cwx & \cwx \qw &  \qw  & \qw & \qw   \\
& \lstick{(S_{11})} & \lstick{8} \ex[1] & \qw & \qw & \qw & \qw & \qw & \targ & \targ & \targ & \meter &  &  &  & \cwx & \cwx & \cwx &  &  & \\
& \lstick{(T)~} & \lstick{11} & \qw \qw & \qw & \qw & \qw & \qw & \qw & \qw & \qw & \targ \cwx & \ctrl{1}  & \qw  & \gate{H} & \meter \cwx & \cw \cwx & \cwx \cw & & &  \\
& \lstick{(T)~} & \lstick{13} &\qw & \qw & \qw & \qw & \qw & \qw  & \qw & \qw & \qw & \gate{\mathcal{H}} & \gate{\mathcal{H}}  & \qw & \qw & \qw & \qw & \qw & \qw & \qw \\
& \lstick{(T)~} & \lstick{12} \ex[1] & \qw & \qw  & \targ & \qw & \qw & \qw & \qw & \qw & \qw & \qw & \ctrl{-1} & \gate{H} & \meter  &  &  &  &  &  \\
& \lstick{(S_{12})~} & \lstick{10} & \qw \qw &  \targ & \meter \cwx & & & & & & & & & & \cwx &  &  &  & &  \\
& \lstick{(S_{12})~} & \lstick{9} & \qw & \ctrl{-1} & \qw & \qw & \qw & \qw & \qw & \qw & \qw & \qw & \qw & \qw & \gate{\sigma_{z}} \cwx & \qw & \qw & \qw & \qw & \qw  
}
 \caption{{\it Simultaneous Control-Hermitian gate implementation through rooted-tree network}. Here the relevant party possesses a qubit given in first bracket. $H$ and $\mathcal{H}$ denote Hadamard and Hermitian gates respectively; the dotted line represents entangled state of two qubit. All the measurements shown here are in computational basis.}
\end{figure*}
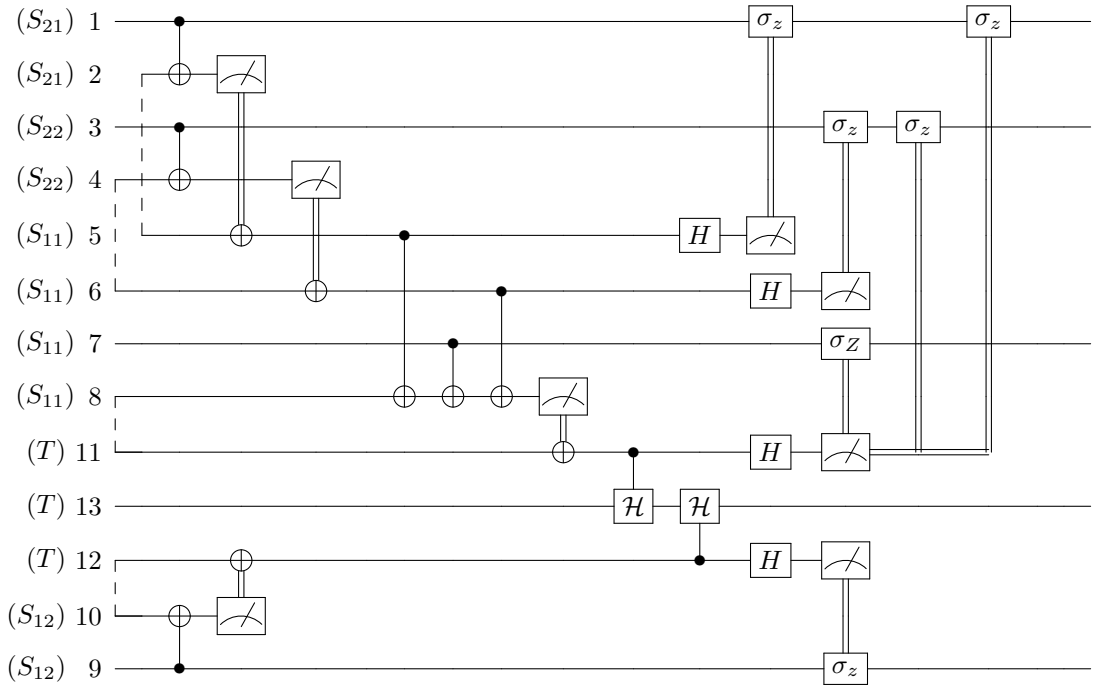

\subsection {Protocol for $n$-party simultaneous implementation of control-Hermitian gate}
We now generalize the above protocol for $n$ remote parties, sharing an unknown $n$-qubit state $|\psi\rangle$ and maintaining an arbitrary rooted-tree entangled network. At first, all the control parties, located at leaves, apply a control-not $(CN)$ gate on their respective pair of qubits taking the target qubit as the entangled qubit shared with another control party at the successive upper depth. In the next step, all the leaves, which are at the maximum depth $(d=h)$ measure their target qubit in computational basis and convey the outcomes to the successive control party located at $d=(h-1)$. Based on the outcome of measurements, these control parties at $d=(h-1)$ depth perform local operations, {\it i.e.}, if the outcome is $ | 1 \rangle $ they apply  $CN^q_{p} CN^{q}_{l}  ~ \sigma^{p}_x$ otherwise $CN^q_{p} CN^{q}_{l}$ for every $p$ and $q$, where $p$ is the entangled qubit shared with the control party at $d=h$ depth and $q$ is the entangled qubit with the qubit of control party at $d=(h-2)$ depth. $l$ is the unknown qubit of the state $|\psi\rangle$ possessed by that party. These control parties then measure their target qubit $q$ in computational basis and convey the outcomes to the successive control party at $d=(h-2)$. Similarly on the basis of outcomes, they perform local operations, measure their target qubit in computational basis and convey to the next control party. This process continues till the control parties at $d=1$ depth is reached. The control parties at depth $d=1$ measure their shared qubit and convey the outcomes to the target party $T$. If the outcome is $|1\rangle$, then $T$ performs local operations $C\mathcal{H}^l_k ~ \sigma^{k}_x$, otherwise only $C\mathcal{H}^l_k$ for every $k$, where $k$ is the entangled qubit shared with the control party at $d=1$ depth and $l$ is the unknown qubit of the type $|\psi\rangle$ possessed by the target party. Subsequently, all the entangled qubits shared by the target party are measured in Hadamard basis and the outcomes of the measurements are conveyed to all the control parties, starting from the respective control party with whom the entangled state has been shared. The other control parties, except the ones in the leaves, measure the shared entangled qubit with the control party at successive lower depth in the Hadamard basis and convey the outcomes to them. Finally all the control parties perform $\sigma_z$ on their possessed qubit of the state $|\psi\rangle$, if odd number of parties (including all the control parties with higher depth and the target party) get the measurement outcome $|1\rangle$.  

Thus, for $n$-party simultaneous implementation of a $C\mathcal{H}$ gate, the total number of cbits required is,
\begin{equation}
\sum_{d=1}^{h} n_{d} + \sum_{d=1}^{h} d.n_{d} = \sum_{d=1}^{m} n_{d}(d+1),
\end{equation}
and the number of implementation steps is $3h+4$.

\section {Implementation of multiparty control-unitary gate}
We now proceed to describe the implementation of multiparty control-unitary gate on the rooted-tree network. The multiparty control-unitary is defined as a generalized form of a Toffoli gate for multi-qubit system, with one target party, where an arbitrary unitary gate acts on the target qubit, only if all the control qubits shared by the remote parties are $|1\rangle$. Thus, given an arbitrary quantum state $|\psi \rangle_{12...n}$, shared by $n$ remote parties, where $n$-th qubit is the target, the desired state which will be realized by LOCC, is given by, $CU^n_{1,2,...,(n-1)} |\psi \rangle_{12...n}$. Here $CU^t_{i,j,k}$ is denoted as the multiparty control-unitary gate, where $i,j,k$ are the control qubits and $t$ is the target. \\

To demonstrate the implementation, first we consider the same five party scenario as shown in Fig. 2. The qubit distribution is described in the earlier section, where the four control parties want to implement a controlled-unitary gate $CU^{13}_{1, 3, 7, 9}$ on an unknown state $|\psi\rangle_{1, 3, 7, 9, 13}$. The protocol for such implementation is described below:\\

{\bf Step 1:} $S_{21}$, $S_{22}$ and $S_{12}$ respectively apply $CN^2_{1}$, $CN^4_3$, $CN^{10}_{9}$,\\

{\bf Step 2:} Then $S_{21}$ and $S_{22}$ measure qubits 2 and 4 respectively in computational basis,\\

{\bf Step 3:} Following that, they convey the outcomes, using classical communication resource to $S_{11}$,\\

{\bf Step 4:} Now, $S_{11}$ applies the following local operations as described in Table \ref{table 4}, depending on the measurement outcomes:\\

\begin{table}[h]
\begin{center}
\begin{tabular}{|c|c|}
\hline 
{\bf Outcome of measurements}& {\bf Local operation after measurements} \\ 
\hline 
$ | 0 \rangle_{2} | 0 \rangle_{4}$ & $CN^8_{5,6,7}$ \\ 
\hline 
$| 1 \rangle_{2} | 0 \rangle_{4}$ & $CN^8_{5,6,7}  ~ \sigma^{5}_x$ \\ 
\hline 
$| 0 \rangle_{2} | 1 \rangle_{4}$ & $CN^8_{5,6,7}  ~ \sigma^{6}_x$ \\ 
\hline
$| 1 \rangle_{2} | 1 \rangle_{4}$ & $CN^8_{5,6,7}  ~ \sigma^{5}_x  ~ \sigma^{6}_x $ \\
\hline 
\end{tabular}
\caption{Operations performed by $S_{11}$ depending upon the outcomes of measurement of qubits 2 and 4 measured by $S_{21}$ and $S_{22}$}
\label{table 4}
\end{center}
\end{table}

{\bf Step 5:} $S_{11}$ and $S_{12}$ measure qubits 8 and 10 respectively in the computational basis.\\

{\bf Step 6:} They convey the outcomes to $T$.\\

{\bf Step 7:} $T$ applies the following unitary operation as shown in Table \ref{table 5}, where $U$ is the multiparty gate, \\

\begin{table}[h]
\begin{center}
\begin{tabular}{|c|c|}
\hline 
{\bf Outcome of measurements}& {\bf Local operation after measurements} \\ 
\hline 
$ | 0 \rangle_{8} | 0 \rangle_{10}$ & $CU^{13}_{11,12} $ \\ 
\hline 
$| 1 \rangle_{8} | 0 \rangle_{10}$ & $CU^{13}_{11,12}   ~ \sigma^{11}_x$ \\ 
\hline 
$| 0 \rangle_{8} | 1 \rangle_{10}$ & $CU^{13}_{11,12}   ~ \sigma^{12}_x$ \\ 
\hline
$| 1 \rangle_{8} | 1 \rangle_{10}$ & $CU^{13}_{11,12}   ~ \sigma^{11}_x  ~ \sigma^{12}_x $ \\
\hline 
\end{tabular}
\caption{Local operations performed by $T$ which depends on the outcome of the measurement of qubit 8 and 10 measured by $S_{11}$ and $S_{12}$ respectively.}
\label{table 5}
\end{center}
\end{table}
{\bf Step 8:} $T$ measures qubits 11 and 12 in Hadamard basis.\\

{\bf Step 9:} Next $T$ conveys the outcomes of qubits 11 and 12 to $S_{11}$ and $S_{12}$ respectively.\\

{\bf Step 10:} $S_{11}$ and $S_{12}$ apply following gates, as shown in Table \ref{table 6}, according to the outcomes ($CZ$ stands for control-$\sigma_z$ gate), \\

\begin{table}[h]
\begin{center}
\begin{tabular}{|c|c|}
\hline 
{\bf Outcome of measurements}& {\bf Local operation after measurements} \\ 
\hline 
$ | + \rangle_{11} | + \rangle_{12}$ & $ \sigma^{9}_z $ \\ 
\hline 
$| + \rangle_{11} | - \rangle_{12}$ & $I$ \\ 
\hline 
$| - \rangle_{11} | + \rangle_{12}$ & $CZ^7_{5,6}   ~ \sigma^{9}_z$ \\ 
\hline
$| - \rangle_{11} | - \rangle_{12}$ & $CZ^7_{5,6} $ \\
\hline 
\end{tabular}
\caption{Operations performed by $S_{11}$ and $S_{12}$ according to the outcomes conveyed by $T$.}
\label{table 6}
\end{center}
\end{table}
 
{\bf Step 11:} After that, $S_{11}$ measures qubits 5 and 6 in Hadamard basis.\\

{\bf Step 12:} Next, $S_{11}$ conveys the outcomes of qubits 6 and 5 mesurements to $S_{21}$ and $S_{22}$ respectively.\\

{\bf Step 13:} Finally $S_{21}$ and $S_{22}$ apply gates to obtain the desired state as described in Table \ref{table 7} : \\

\begin{table}[h]
\begin{center}
\begin{tabular}{|c|c|}
\hline 
{\bf Outcome of measurements}& {\bf Local operation after measurements} \\ 
\hline 
$ | + \rangle_{5} | + \rangle_{6}$ & $ \sigma^{3}_z $ \\ 
\hline 
$| + \rangle_{5} | - \rangle_{6}$ & $I$ \\ 
\hline 
$| - \rangle_{5} | + \rangle_{6}$ & $  \sigma^{1}_z  ~ \sigma^{3}_z$ \\ 
\hline
$| - \rangle_{5} | - \rangle_{6}$ & $ \sigma^{1}_z $ \\
\hline 
\end{tabular}
\caption{Operations performed by $S_{21}$ and $S_{22}$ to obtain the desired state, depending on the outcome conveyed by $S_{11}$}
\label{table 7}
\end{center}
\end{table}

The pictorial representation of the above protocol is shown in Fig.4. The number of ebits and cbits required to implement a control-unitary gate are 4 and 8 respectively. \\

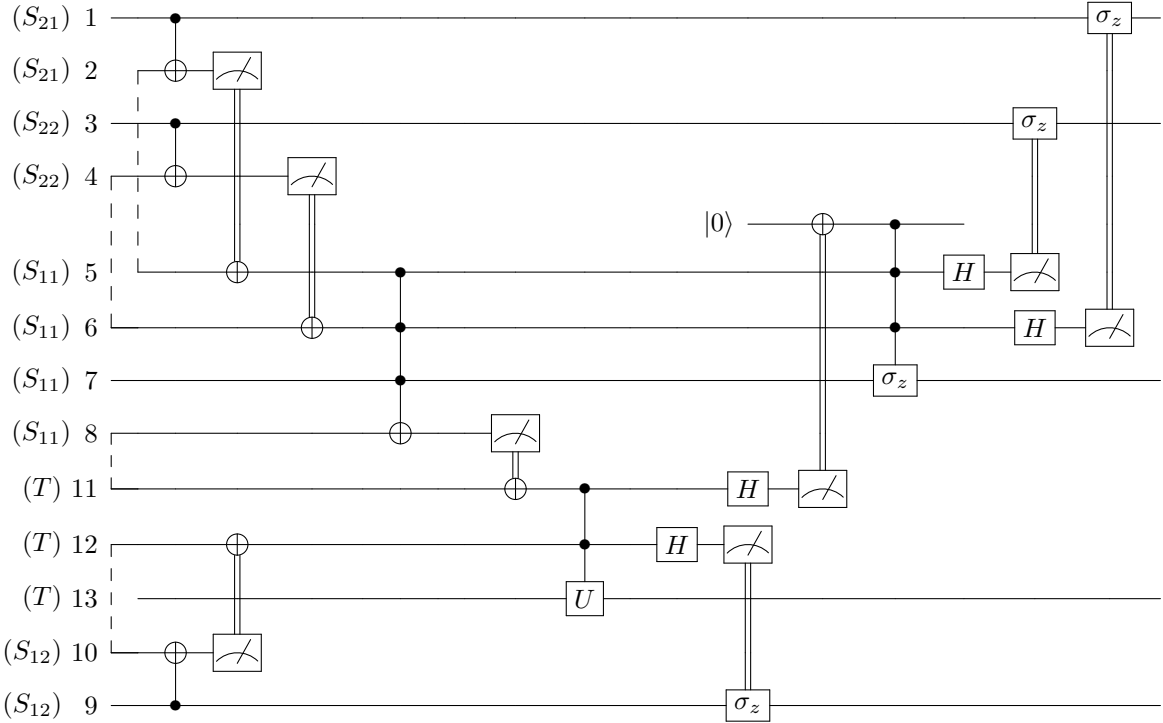
\begin{figure*}[h]
   \centering
\Qcircuit @C=1em @R=.7em
{
 & \lstick{(S_{21})} & \lstick{1} & \qw & \ctrl{1} & \qw & \qw & \qw & \qw & \qw & \qw  & \qw & \qw & \qw  & \qw & \qw & \qw & \qw & \qw & \qw & \gate{\sigma_{z}} & \qw  \\
& \lstick{(S_{21})} & \lstick{2} & \ex[4] & \targ & \meter &  &  &  &  &  &  &  &  &   &  & &  &  &  &  \cwx &   \\
& \lstick{(S_{22})} & \lstick{3} & \qw  & \ctrl{1} & \qw \cwx & \qw & \qw & \qw & \qw & \qw & \qw & \qw & \qw  & \qw  & \qw & \qw & \qw  & \qw & \gate{\sigma_{z}}  & \qw \cwx & \qw \\
& \lstick{(S_{22})} & \lstick{4} \ex[3]  & \qw  & \targ & \qw \cwx & \meter &  &  &  &  &  &  & & &  & & &  &  \cwx  &  \cwx &  \\
& & & & & \cwx & \cwx & & & & & & & & & \lstick{| 0 \rangle} & \targ & \ctrl{1} & \qw & \cwx & \cwx \\
& \lstick{(S_{11})} & \lstick{5} & ~ & \qw & \targ \cwx & \qw \cwx & \qw & \ctrl{1} & \qw & \qw & \qw  & \qw  & \qw  & \qw  & \qw  & \qw \cwx & \ctrl{1}  &  \gate{H} & \meter \cwx & \cwx &  \\
& \lstick{(S_{11})} & \lstick{6} & \qw \qw & \qw & \qw & \targ \cwx & \qw & \ctrl{1} & \qw & \qw & \qw  & \qw & \qw  & \qw  & \qw  & \qw \cwx & \ctrl{1}  & \qw & \gate{H} & \meter \cwx &   \\
& \lstick{(S_{11})} & \lstick{7} & \qw & \qw & \qw & \qw & \qw  & \ctrl{1} & \qw & \qw & \qw & \qw & \qw  & \qw & \qw & \qw \cwx & \gate{\sigma_{z}} & \qw  & \qw & \qw & \qw   \\
& \lstick{(S_{11})} & \lstick{8} \ex[1] & \qw  & \qw & \qw & \qw & \qw & \targ & \qw & \qw & \meter & & &  & & \cwx & &  &  &  &  \\
& \lstick{(T)~} & \lstick{11} & \qw \qw & \qw & \qw & \qw & \qw & \qw & \qw & \qw & \targ \cwx & \ctrl{1} & \qw & \qw & \gate{H} & \meter \cwx &  &  &  &  &   \\
& \lstick{(T)~} & \lstick{12} \ex[2] & \qw & \qw  & \targ & \qw & \qw & \qw & \qw & \qw & \qw  & \ctrl{1} & \qw & \gate{H} & \meter &  &  &  &  &  &   \\ 
& \lstick{(T)~} & \lstick{13} & ~ & \qw & \qw \cwx & \qw & \qw & \qw  & \qw & \qw & \qw & \gate{U} & \qw  & \qw & \qw \cwx & \qw & \qw & \qw & \qw & \qw & \qw \\
& \lstick{(S_{12})~} & \lstick{10} & \qw \qw &  \targ & \meter \cwx & &  &  &  & &  &  & & &  \cwx &  &  &   &  &  &   \\
& \lstick{(S_{12})~} & \lstick{9} & \qw & \ctrl{-1} & \qw & \qw & \qw & \qw & \qw & \qw & \qw & \qw & \qw & \qw & \gate{\sigma_{z}} \cwx & \qw & \qw & \qw & \qw & \qw & \qw   
}
\caption{\it multiparty control-unitary gate implementation through rooted tree network}
\end{figure*}

\subsection {Protocol for implementation of $n$-party control-unitary gate}
The generation of the above protocol, for the $n$ qubit unknown state $|\psi\rangle$ in an arbitrary rooted-tree network is as follows. At first,
 all the control parties which are located on the leaf apply a control-not $(CN)$ gate on their qubits, taking the shared Bell state as the target qubit. In the next step, all the parties which are at maximum depth $(d=h)$ measure their target qubit in computational basis and convey the outcome to the control party at $d=(h-1)$ depth. Based on the outcomes of the measurements, these control parties at $d=(h-1)$ depth perform local operation $CN^{q}_{p,l}  ~ \sigma^{p}_x$, when the outcome is $ | 1 \rangle $, otherwise $CN^{q}_{p,l}$ for all $p$ and $q$, where $p$ is the entangled qubit shared with the control party at $d=h$ depth, and $q$ is the entangled qubit shared with the control party at $d=(h-2)$ depth. $l$ is the qubit of the state $|\psi\rangle$ possessed by that party. 
All these control parties then measure their target qubit $q$ in computational basis and convey the outcomes to the respective control party at $d=(h-2)$ depth. This process continues upto the control parties at $d=1$ depth, where the parties measure their entangled qubit and convey the outcomes to the target party. 
$T$ performs local operations $\sigma^{i}_x$, only if the outcome is $|1\rangle$ for every $i$, where $i$ is the entangled qubit shared with the control party at $d=1$ depth. Then $T$ applies $C\mathcal{H}^l_{i,j,...}$, where $l$ is the unknown qubit of the state $|\psi\rangle$ possessed by the target party and $i,j...$ are entangled qubits shared with each of the control party at $d=1$. After that, all the entangled qubits shared by the target party are measured in the Hadamard basis and the outcomes of the measurement is conveyed to all the control parties at depth $d=1$. These control parties at $d=1$, apply $CZ^l_{a,b,...}$, only if the outcomes in the Bell shared qubit with target party is $|1\rangle$. Here $a,b,..$ are entangled qubits shared with the control parties at $d=2$ and $l$ is the qubit of the state $|\psi\rangle$ possessed by that party. Then they measure $a,b,...$ qubits in Hadamard basis and convey the outcomes to respective control parties at lower depth. This continues upto the leaves, where they perform $\sigma_z$ operation if the outcome of the shared entangled qubit by the control party in the successive upper depth is $|1\rangle$. This completes the protocol.

For implementation of $n$-party control unitary gate, the total number of implementation steps is $6h+1$ and the number of cbits required is,
\begin{equation}
2\sum_{d=1}^{h} n_{d} = 2(n-1)
\end{equation}

The comparison of gate teleportation in the present rooted-tree network and the earlier networks are summarized in the Tables \ref{table 8} and \ref{table 9} given below,

\begin{table}[h]
\begin{center}

\begin{tabular}{|c|c|c|}
\hline 
{\bf Network channel}  & {\bf Classical communication cost} & {\bf No. of steps for implementation} \\ 
\hline 
Parallel  & $2(n-1)$ & 7\\ 
\hline 
Linear   & $\frac{n^2 + n -2}{2}$ & $3n+1$\\ 
\hline 
Rooted-tree & $\sum_{d=1}^{h} n_{d}(d+1)$ & $3h+4$\\ 
\hline 
\end{tabular}
\caption{Comparison of classical communication cost and number of steps for implementing simultaneous control-Hermitian gate on different entangled channels, $n$ is total number of remote parties and $h$ is the height of the rooted tree}
\label{table 8}
\end{center}
\end{table}

\begin{table}[h]
\begin{center}
\begin{tabular}{|c|c|c|}
\hline 
{\bf Network channel} & {\bf Classical communication cost} & {\bf No. of steps for implementation} \\ 
\hline 
Parallel  & $2(n-1)$ & 7 \\ 
\hline 
Linear  & $2(n-1)$ &  $6n-5$\\ 
\hline 
Rooted-tree  & $2(n-1)$ & $6h+1$\\ 
\hline 
\end{tabular} 
\caption{Comparison of classical communication cost and number of steps for implementing multiparty control-Unitary gate on different entangled channels}
\label{table 9}
\end{center}
\end{table}

\section {Discussion }
We have shown that a rooted-tree network is the most general entanglement distribution, which provides a fresh perspective for implementing non-local gates, locally from several control parties to a target party, under the condition of optimal entanglement cost. The explicit protocols of simultaneous implementation of Controlled-Hermitian gate and the multiparty Controlled-Unitary gate are described here in detail.\\
The earlier explored entangled networks, namely $parallel$ and $linear$ configurations are specific cases of the rooted-tree network, where $h=1$ and $h=n-1$ respectively. Its worth mentioning that, for all the cases, the entanglement cost is $(n-1)$ ebits, which is optimal. It can be seen that, for a given $n$, the classical communication cost and the number of steps for implementing simultaneous control-Hermitian gate increases as we go from parallel to linear network, through an arbitrary rooted-tree structure. For multiparty control-unitary gate, only the number of implementation steps increases. Significantly, the number of entangled states maintained by a party decreases as the height of rooted-tree increases. For example, in the scenario consider in Fig. 2, the maximum number of entangled  states maintained by a party is 3; it is 4 for parallel and 2 for linear network. Thus the novelty of the rooted-tree entanglement distribution between $n$ remote parties lies in the fact that, it opens up the freedom of designing the entanglement network according to the capacity of maintaining entangled channels of each party, as well as the suitable number of implementation steps and the available classical communication resources. \\
In future, it will be of interest to investigate whether the classical communication cost for simultaneous control-Hermitian gate teleportation described here is optimal or not. Teleportation of other non-local gates using local operations and classical communication is also worth studying.
\paragraph*{Acknowledgement:}D.S. acknowledges support from NCN grant 2013/08/M/ST2/00626. N.V. acknowledges Inspire programme undertaken by Department of Science and Technology, Government of India, for support.

\end{document}